\documentclass[12pt]{article}
\usepackage{amssymb,amsmath,amscd,amsthm}
\usepackage[utf8x]{inputenc}
\usepackage[T2A]{fontenc}

\usepackage{graphicx,psfrag}

\newtheorem{theorem}{Theorem}[section]

\newtheorem{lemma}[theorem]{Lemma}

\newcommand{\sgn}{\text{sgn}}

\date{}

\begin{document}
 \author{ Evgeny Lakshtanov\thanks{Department of Mathematics, Aveiro University, Aveiro 3810, Portugal.  This work was supported by Portuguese funds through the CIDMA - Center for Research and Development in Mathematics and Applications and the Portuguese Foundation for Science and Technology (``FCT--Fund\c{c}\~{a}o para a Ci\^{e}ncia e a Tecnologia''), within project UID/MAT/0416/2013 (lakshtanov@ua.pt)} 
 \and
 Boris Vainberg\thanks{Department
of Mathematics and Statistics, University of North Carolina,
Charlotte, NC 28223, USA. The work was partially supported  by the NSF grant DMS-1410547 (brvainbe@uncc.edu).}}

\title{ Recovery of $L^p$-potential in the plane}

\maketitle

\begin{abstract}
An inverse problem for the two-dimensional Schrodinger equation with $L^p_{com}$-potential, $p>1$, is considered. Using the $\overline{\partial}$-method, the potential is recovered from the Dirichlet-to-Neumann map on the boundary of a domain containing the support of the potential. We do not assume that the potential is small or that the Faddeev scattering problem does not have exceptional points.  The paper contains a new estimate on the Faddeev Green function that immediately implies the absence of exceptional points near the origin and infinity when $v\in L^p_{com}$.
\end{abstract}

\textbf{Key words:}
$\overline{\partial}$-method, Faddeev scattering problem, inverse scattering problem.

\section{Introduction}
The article is devoted to the reconstruction of the potential $v(z)$ in a bounded domain $\mathcal O$ using the Dirichlet-to-Neumann map $\Lambda_v$ (defined in Appendix, part V) on $\partial\mathcal O$ for the equation\footnote{There is an inconsistency in the choice of the sign for the potential $v$ in the journal version of this paper. One needs to replace $v$ by $-v$ in formulas (1), (2), (4) and in the statement of Theorem 2.3, and replace $S_\lambda$ by $-S_\lambda$ in (6) and in the formula after (6) in the journal version. Other corrections that need to be made are very minor and most of them are in  Appendix V.}
\begin{equation}\label{fir}
(-\Delta-E+v)u(x) =0,~~~  x \in \mathcal O\subset \mathbb R^2,
\end{equation}
with a fixed positive value of energy $E$. We extend the potential by zero in $\mathbb R^2\backslash\mathcal O$ and assume that it belongs to the space $ L_{com}^{p},~p>1,$ of functions in $L^p(\mathbb R^2)$ with support in $\overline{\mathcal O}$. One can choose an arbitrary value of energy $E$ by changing $v$ in $\mathcal O$ by a constant, since  $v$ is not assumed to be smooth in $\mathbb R^2$. We assume only that $E$ is not an eigenvalue of the Dirichlet problem for the operators $-\Delta+v$ and $-\Delta$ in $\mathcal O$ in order to be able to consider the Dirichlet-to-Neumann maps on $\partial\mathcal O$ for these operators (this condition on $E$ can be satisfied by enlarging $\mathcal O$ slightly, if needed). One can also assume that the boundary $\partial\mathcal O$ is (infinitely) smooth since the domain $\mathcal O$ can be enlarged.

We consider the non-over-determined two-dimensional problem, and we do not assume that the potential is small. The main definitions will be given for complex-valued potentials since they can be treated by the same technique as the real-valued ones. However,  the results will be proved only for real-valued $v$ since the complex-valued case requires additional smoothness assumptions.  We will not distinguish functions of complex variable $z=x_1+ix_2$ and functions of two real variables $x=(x_1,x_2)$.

One of the most common approaches (called the $\overline{\partial}$-method) to the reconstruction of the potential (e.g., \cite{nov86}, \cite{Novikov86}, \cite{gm}, \cite{RN2}, \cite{nachman}) is based on the Faddeev scattering solution
$$
\psi(\lambda,z,E), \quad \lambda\in \mathbb C'=\mathbb C\backslash(\{0\}\bigcup\{|\lambda|=1\}
$$
and the well known $\overline{\partial}$-equation for $\psi$. Function
$\psi$ is the solution of the problem
 \begin{equation}\label{scsol}
  (-\Delta-E+v)\psi =0,~~~  x \in \mathbb R^2,
 \quad   \psi e^{-i\frac{\sqrt{E}}{2} (\lambda \overline{z} + z / \lambda) }\to 1, ~~ |z|\to\infty.
\end{equation}
The exponent above grows or decays at infinity depending on the direction. So, we will avoid a discussion of the unique solvability of (\ref{scsol}) by using an integral form of the problem above. The unique solvability of the corresponding integral (Lippmann-Schwinger) equation may be violated for certain $\lambda\in \mathbb C'$. These points are called exceptional.

The scattering data $h(\varsigma,\lambda)$ of the Faddeev scattering problem is defined by the following formula involving the Cauchy data of $\psi(\lambda,z)$  on $\partial\mathcal O$:
\begin{equation}\label{hh}
  h(\varsigma,\lambda)=\frac{1}{(2\pi)^2} \int_{\partial\mathcal O}[ e^{-i\frac{\sqrt{E}}{2} (\varsigma \overline{z}+z / \varsigma)}\frac{\partial}{\partial\nu}\psi(z,\lambda) - \psi(z,\lambda)\frac{\partial}{\partial\nu} e^{-i\frac{\sqrt{E}}{2} (\varsigma \overline{z}+z / \varsigma)}]dl_z, \quad \varsigma,\lambda\in \mathbb C',
   \end{equation}
where $\nu $ is the outer unit normal to $\partial\mathcal O$.

Let us stress that we do not plan to solve (\ref{scsol}) since the potential $v$ is unknown. However, the Dirichlet-to-Neumann map $\Lambda_v$ allows one to determine easily the Cauchy data of $\psi$ and find the scattering data $h$. We will show that $\psi$ can be recovered if $h$ is known. Then the potential $v$ can be found, for example from (\ref{scsol}):
\begin{equation}\label{vav}
 v(z)=\frac{(\Delta+E)\psi}{ \psi}.
\end{equation}
Thus the $\overline{\partial}$-method can be described by the following diagram:
\begin{equation}\label{dia}
\Lambda_v\to(\psi|_{\partial\mathcal O},\frac{\partial\psi}{\partial \nu}|_{\partial\mathcal O})\to h\to\psi\to v,
\end{equation}
with the most difficult step being between $h$ and $\psi$. Let us provide the exact formulas for the diagram above.

The first step is well known. Function $\psi|_{\partial\mathcal O}$ can be determined by $\Lambda_v$ as follows
\begin{equation}\label{dtnpsi}
\psi(z,\lambda)=(I-S_\lambda(\Lambda_{v}-\Lambda_0))^{-1}e^{i\frac{\sqrt{E}}{2} (\lambda \overline{z} + z / \lambda) }, \quad z \in \partial \mathcal O,
\end{equation}
where $S_\lambda$ is the single layer operator corresponding to Faddeev's Green function (\ref{2DecA}). This formula appeared first in \cite[(5.18)]{novkhen}. For $v\in L_p,p>1,$ it has been justified in \cite[Th.5]{nachman} when $E=0$, but the proof remains the same if $E>0$. Formula (\ref{dtnpsi})  in a different form
$$
(I-S_\lambda(\Lambda_{v}-\Lambda_0))\psi(z,\lambda)=e^{i\frac{\sqrt{E}}{2} (\lambda \overline{z} + z / \lambda) }, \quad z \in \partial \mathcal O, 
$$
follows immediately from the Green formula. It is shown in \cite{novkhen}, \cite{nachman} that the operator on the left-hand side above is invertible if and only if the Faddeev scattering solution $\psi$ exists. Thus (\ref{dtnpsi}) holds for all $\lambda$ for which $\psi$ exists.

Let us specify (\ref{dtnpsi}) a little bit more. We assumed that $\partial\mathcal O$ is smooth, and therefore we do not need to consider weak solutions of (\ref{fir}) as it is done in \cite{nachman}. It will be shown in Appendix, part V,  that equation (\ref{fir}) with the boundary condition $u_0:= u|_{\partial\mathcal O}\in W^{2-1/p,p}(\partial \mathcal O)$ is uniquely solvable in $W^{2,p}(\partial \mathcal O)$ (see the definitions of above spaces in e.g., \cite{agdn}, \cite{agranovich}). Then
\[
\Lambda_{v},\Lambda_0:W^{2-1/p,p}(\partial \mathcal O)\to W^{1-1/p,p}(\partial \mathcal O)
 \]
 are bounded operators, and $S_\lambda(\Lambda_{v}-\Lambda_0)$ is a compact operator in $W^{2-1/p,p}(\partial \mathcal O)$. Hence  (\ref{dtnpsi}) can be considered as a relation in $W^{2-1/p,p}(\partial \mathcal O)$.

One also can avoid working with spaces $W^{s,p}$. We may assume that $v=0$ in a neighborhood of $\partial \mathcal O$ (recall that $\mathcal O$ can be enlarged). Then from local a priori estimates for solutions of elliptic problems it follows that the solution $u\in W^{2,p}(\partial \mathcal O)$ of (\ref{fir}) constructed in Appendix, part V, can be assumed to be sufficiently smooth near  $\partial \mathcal O$ if the boundary condition is smooth enough. Similarly, $\psi\in  C^\infty$ in a neighborhood of $\partial \mathcal O$. The operator $S_\lambda(\Lambda_{v}-\Lambda_0)$ is infinitely smoothing in this case and compact in all the Sobolev spaces, i.e., (\ref{dtnpsi}) can be considered as a relation in $H^s(\partial \mathcal O)$ with an arbitrary $s$.

After $\psi|_{\partial \mathcal O}$ is found (by (\ref{dtnpsi})), one can evaluate $\frac{\partial\psi}{\partial\nu}|_{\partial \mathcal O}=\Lambda_v\psi|_{\partial \mathcal O}$ and find $h$ using (\ref{hh}). Since $\psi$ determines the potential $v$ via (\ref{vav}), it remains to describe only one step in the diagram (\ref{dia}): the transition from $h$ to $\psi$.

Function $\psi$ could be determine from the following $\overline{\partial}$-equation (to be derived in Appendix, part III):
\begin{equation}\label{dbar2}
\frac{\partial}{\partial \overline{\lambda}} \psi (z,\lambda)=r(\lambda) \psi\left (z,-\frac{1}{\overline \lambda} \right ), \quad |\lambda| \neq 0,1,
\end{equation}
complemented by specific asymptotic behavior at infinity. Function $r$ here is defined by $h$:
 \[
    r(\lambda)=  \frac{\sgn(|\lambda|^2-1)\pi}{\overline{\lambda}} h(-\frac{1}{\overline{\lambda}},\lambda).
\]
However, this approach requires the existence of $\psi $ and the validity of (\ref{dbar2}) for all the values of $\lambda$. So, until recently, the $\overline{\partial}$-method was restricted by the assumption of the absence of {\it exceptional points}, that is the points $\lambda\in \mathbb C'$ for which the unique solvability of the Faddeev scattering problem is violated. We will show that $\psi$ can be found in the presence of exceptional points (or curves) by the following procedure.

We will show that the set of exceptional points is separated from zero and infinity, i.e., there exists a ring $D=\{  \lambda \in \mathbb C: A^{-1}<|\lambda| < A\}$ that contains all the exceptional points, and $\psi$ exists when $\lambda \in \mathbb C\backslash D$ (see
Lemma \ref{l21}). Consider the space
\[
\mathcal H^s :=  L^s(\mathbb R^2) \cap C(\overline{D}) , \quad s>2,
\]
and the operator $T_z=T_z^{(1)} +T_z^{(2)}:\mathcal H^s\to \mathcal H^s$, where
\begin{equation}\nonumber
T_z^{(1)} \phi=
-\frac{1}{\pi} \int_{\mathbb C \backslash D}  r(\varsigma) e^{-i\frac{\sqrt{E}}{2} (\varsigma\overline{z}+z / \varsigma +\frac{1}{\overline \varsigma} \overline{z}+z \overline{\varsigma})} \phi(\frac{-1}{\overline \varsigma})   \frac{d\varsigma_R d\varsigma_I}{\varsigma-\lambda} ,
\end{equation}
\begin{equation}\nonumber
T_z^{(2)} \phi=\frac{1}{2\pi i} \int_{\partial D} \frac{d\varsigma}{\varsigma - \lambda} \int_{\partial D} c(\varsigma,\varsigma')h(\varsigma',\varsigma) e^{-i\frac{\sqrt{E}}{2} (\varsigma \overline{z} + z / \varsigma) }e^{i\frac{\sqrt{E}}{2} (\varsigma' \overline{z} + z / \varsigma') }\phi^-(\varsigma')d\varsigma'.
\end{equation}
Here $\phi^-$ is the  trace of $\phi$ on $\partial D$ taken from the interior of $D$, function $c(\varsigma,\varsigma')$ is given in~(\ref{4Jan1}). Operator $T_z$ acts on functions of $\lambda\in \mathbb C$, and the space variable $z \in \mathbb C$ serves as a parameter.

We will show that operator $T_z$ is compact and depends continuously on $z\in \mathcal C$. Moreover, if $s > \widetilde{s}:=\max(q,4),~p^{-1}+q^{-1}=1,$ then for every $z_0\in \mathbb C$ and a generic  potential $v\in L^{p}_{com}$, $p>1$, the equation
  \begin{equation}\label{isp1}
 (I+T_z)(\phi(z,\cdot)-1) =-T_z1, \quad \phi -1 \in \mathcal H^s, \quad
 \end{equation}
  is uniquely solvable for all $z$ in some neighborhood of $z_0$ in $\mathbb C$. For each $\lambda\in\mathbb C\backslash D$, function $\psi=e^{i\frac{\sqrt{E}}{2} (\lambda \overline{z} + z / \lambda) }\phi(z,\lambda)$  is the Faddeev scattering solution and defines $v$ by (\ref{vav}) when $|\lambda|+\frac{1}{|\lambda|}$ is large enough. The latter restriction on $|\lambda|$ is needed to guarantee that $\psi\neq 0$ and the denominator in (\ref{vav}) does not vanish.

There is an alternative way to recover the potential using the same scattering solution $\psi$ in 3D (see \cite[Th.1.1]{novkhen}) or another parameter-dependent  family of non-physical scattering solutions and their scattering data in 2D (the Bukhgeim  approach, see e.g., \cite{b}, \cite{afr}, \cite{tejero}). Then a connection is established between $v$ and the asymptotic behavior of the scattering data (or their analogue) as the parameter goes to infinity. After that, $v$ is recovered from the asymptotic behavior of the scattering data at infinity without the knowledge of $\psi$.

One of the advantages of the alternative approach is its generality: the method does not require the existence of $\psi $ for all values of $\lambda$. On the other hand, the $\overline{\partial}$-method is more stable. It was shown in \cite{nov2}, \cite{sant}, \cite{blasten}  that the Bukhgeim  approach (based solely on the asymptotics of data at infinity) has a logarithmic stability, while the integral equations approach (that uses $\psi$ for all $\lambda$) allows the reconstruction with stability of a H$\ddot{o}$lder type,  and the  stability increases with the smoothness of the potential. Paper \cite{blasten} justifies the uniqueness for the inverse problem for potentials from $L^p_{com}(\mathbb R^2)$, $p>2$. Let us stress (and this is more important than the issue of stability) that the $\overline{\partial}$-method has its independent value as a tool for solving certain nonlinear equations, see \cite{astala} and the references therein.

The absence of exceptional points has been shown in \cite{isakovnachman} under assumptions that $E=0$ and the Dirichlet problem for the operator $-\Delta+v$ in $\mathcal O$ does not have negative eigenvalues. The paper also contains an implementation of the $\overline{\partial}$-method to recover $v$  for real-valued potentials from $L^p(\mathcal O), ~p>1$, under the above assumptions.  Similar results were obtained in \cite{nachman} for conductivity-type potentials in the case of $E=0$, and this class of potentials was widened a little in \cite{music}.\footnote{ a potential is of conductivity-type if $v=-q^{-\frac{1}{2}}\Delta q^{\frac{1}{2}}$, where $q$ is smooth, non-negative, and $q-1$ vanishes outside $\mathcal O$.} In the absence of exceptional points, the integral equation method (at positive energy) has been implemented numerically in \cite{hoop}.

Recently, a reconstruction method has been proposed \cite{afr} for potentials in $H^{s}(\mathbb R^2), s \geq 1/2,$ using Bukhgeim's approach. The latter paper also contains examples of potentials from $H^{1/2}$ for which Bukhgeim's reconstruction scheme fails. Similar results on reconstruction were obtained in \cite{tejero} for piece-wise smooth potentials.

The main result of the present paper is a general (in a possible presence of exceptional points) $\overline{\partial}$-method for reconstruction of potentials $v$ from $ L^p_{com},~p>1$. We follow our recent paper with R. Novikov \cite{NLV}\footnote{The latter paper contains also a discussion on application of  the $\overline{\partial}$-method to the (non-linear) Novikov-Veselov equation; the focusing Davey-Stewartson II system can be treated similarly, \cite{lv16}.}, where this problem was solved for $v\in L^\infty_{com}$.  The idea of the method is the following.
We consider the $\overline{\partial}$-equation (\ref{dbar2}) in the region $\mathbb C\backslash D$, which does not contain exceptional points. One can not reduce (\ref{dbar2}) in  $\mathbb C\backslash D$ to an equivalent Fredholm integral equation without imposing additional boundary conditions for $\psi$ at $\partial D$. It was shown in \cite{NLV} that the jump  at the boundary of $D$ between the function $\psi$ and a certain known solution $\psi^+$ of the Helmholtz equation in $D$ can be represented as an integral operator acting on $\psi^+|_{\partial D}$.  This relation between $\psi$ and $\psi^+$ plays the role of the boundary condition for $\psi$. A prototype of this idea has been proposed in \cite[Sect.8]{RN2}.
Below we justify this boundary condition by application of the Cauchy-Pompeio formula and not by generalized Cauchy formula from the theory of generalized analytic functions, as it was done in \cite{NLV}.
As a result we reduce equation (\ref{dbar2}) in  $\mathbb C\backslash D$ to (\ref{isp1}).


We would like to provide an exposition that does not require a deep knowledge of $\overline{\partial}$-method or the Faddeev scattering theory. Thus we will prove (or give outlines of the proofs) all the most important results from the latter two topics that will be used in our paper. We still will need to rely only on references in some cases, but only for more technical results.

\section{Main results}
Let us fix an arbitrary $E > 0$. Consider the set of vectors $k=(k_1,k_2) \in \mathbb C^2, k_1^2+k_2^2=E >0 ,  $ and its parametrization with $\lambda \in \mathbb C':=\mathbb C\backslash(\{0\}\bigcup\{|\lambda|=1\})$:
\begin{equation}\label{kla}
k_1=\left(\lambda + \frac{1}{\lambda} \right ) \frac{\sqrt{E}}{2},\quad k_2=\left( \frac{1}{\lambda} - \lambda \right ) \frac{i\sqrt{E}}{2},~~\lambda = \frac{k_1 + ik_2}{\sqrt{E}}.
\end{equation}
Note that if ${x}=(x_1,x_2) \in \mathbb R^2$ and $z=x_1+ix_2$, then
\begin{equation}\label{exp}
e^{ik {x}}= \varphi_0(z,\lambda):=e^{i\frac{\sqrt{E}}{2} (\lambda \overline{z} + z / \lambda) }.
\end{equation}

Consider solutions $\psi$ of the Faddeev scattering problem. The incident waves $e^{ik{x}}$ and the scattered waves in the problem grow exponentially at infinity, and the easiest way to define the solution $\psi$ of the Faddeev scattering problem is by using the Lippmann-Schwinger equation:
\begin{equation}\label{19JanA}
\psi(z,\lambda)=e^{ik  {x}}+ \int_{\mathbb R^2} G(z-z',k(\lambda)) v({z'}) \psi({z'},\lambda) d{x_1'}dx_2',
\end{equation}
where $z=x_1+ix_2\in \mathbb C,~e^{ik  {x}}$ can be rewritten in the form of (\ref{exp}), and $G$ is a specific fundamental solution of the operator $\Delta +E$ of the form
\begin{equation}\label{2DecA}
G(z,k)=g(z,k)e^{ik {x}}, \quad  g(z,k)=- \frac{1}{(2\pi)^2}
\int_{\xi \in \mathbb R^2} \frac{e^{i\xi {x}}}{|\xi|^2 + 2k \cdot \xi} d \xi, \quad  k \in \mathbb C^2, ~\Im k \neq 0.
\end{equation}

Often we will use function  $\mu$ instead of $\psi$:
\begin{equation}\label{mu1}
\mu(z,\lambda)= \psi(z,\lambda) e^{-ik{x}}= \psi(z,\lambda) \varphi_0(z,-\lambda).
\end{equation}
Unlike $\psi$, the latter function does not grow, but approaches one at infinity: $\mu=1+o(1),~|z|\to \infty$. The Lippmann-Schwinger equation (\ref{19JanA}) for function $\mu$ takes the following form:
\begin{equation}\label{lsmu}
\mu(z,\lambda)=1+ \int_{ \mathbb R^2} g(z-{z'},k(\lambda)) v({z'}) \mu({z'},\lambda) d{x_1'}dx_2'.
\end{equation}
All the ingredients of this equation (the integral kernel, the right-hand side and the solution) are bounded, while they are growing exponentially at infinity in (\ref{19JanA}). {\it Exceptional points} are defined as points  $\lambda\in \mathbb C'$ for which equation (\ref{lsmu}) is
not uniquely solvable in  $L^\infty$. For bounded potentials, it is known that there is a finite ring in $\mathbb C$ that contains all the exceptional points. This fact will be proved in Lemma \ref{l21} for arbitrary $v\in L^p_{com},~p>1$.


We will not mark dependence of $\psi$ and other functions on energy $E$, since  $E$ is fixed throughout the paper.

Let us define two functions that are called {\it scattering data}. They are given by the formulas for non-exceptional $\lambda \in \mathbb C'$:
\begin{equation}\label{27dec2}
h(\varsigma,\lambda)  =  \frac{1}{(2\pi)^2} \int_{\mathbb R^2}  e^{-i\frac{\sqrt{E}}{2} (\varsigma \overline{z}+z / \varsigma)}v(z)\psi(z,\lambda)d{x_1}dx_2, ~ \quad \varsigma=\sigma_1+i\sigma_2\in \mathbb C,
\end{equation}
and
\begin{equation} \label{27dec1}
r(\lambda)=  \frac{\sgn(|\lambda|^2-1)\pi}{\overline{\lambda}} h(-\frac{1}{\overline{\lambda}},\lambda).
\end{equation}

To justify the use of the term ``scattering data'', one can use the direct analogy of (\ref{27dec2}) with the formula for the scattering amplitude in the classical scattering (if the exponent in (\ref{27dec2}) is replaced by (\ref{exp})). One also can consider an arbitrary domain $\mathcal O$ containing the support of $v$, replace the region of integration $\mathbb C$ above by $\mathcal O$, replace $v(z)\psi(z,\lambda)$ by $(\Delta+E)\psi$, and apply the Green formula. Then $h$ can be rewritten as in (\ref{hh}).

The following uniform estimate will be proved in the Appendix, part I: for each $\alpha\in (0,1]$ there is $c=c(\alpha)$ such that
\begin{equation}\label{19DecD}
|g(z,k(\lambda)) | \leq\frac{c(\alpha)}{[|z|\sqrt E (|\lambda|+|1\backslash \lambda|)]^\alpha}, \quad z,\lambda \in \mathbb C\backslash \{0\}, |\lambda|\notin \left [\frac{1}{2},2\right].
\end{equation}
When $\alpha=1/2$, the estimate was proved in \cite[Prop.3.1]{RN2},  and this particular case could be sufficient to prove the main result if $v\in L^p,~p>4/3$. We need (\ref{19DecD}) with arbitrary $\alpha\in (0,1]$ to consider potentials $v\in L^p$ with $p>1$. In fact, we will prove a slightly stronger estimate. Let $f(\tau)=\ln\frac{1}{\tau}$ when $0<\tau<1/2$, and $f(\tau)=\frac{1}{\tau}$ when $\tau\geq1/2$. Then
\begin{equation}\label{dddd}
|g(z,k(\lambda)) | \leq Cf(|z|\sqrt E (|\lambda|+|1\backslash \lambda|)), \quad z,\lambda \in \mathbb C\backslash \{0\}, |\lambda|\notin \left [\frac{1}{2},2\right].
\end{equation}
Obviously, (\ref{dddd}) immediately implies (\ref{19DecD}).

Estimate (\ref{19DecD}) implies the absence of the exceptional points in some neighborhoods of $\lambda=0,\infty$. To be more exact, the following statement is valid:
\begin{lemma}\label{l21}
For each $v\in L^p_{com},~p>1,$ there exists a ring
$$
D=\{  \lambda \in \mathbb C: A^{-1}<|\lambda| < A\}
$$
such that there are no exceptional points outside $D$. Here $A$ depends on $p, ~\|v\|$, and the radius of a ball containing the support of $v$.
\end{lemma}

{\bf Proof.} The Hölder  inequality implies that
\[
|g*(v\mu)|=|\int_{\mathcal O}g(z-z',k(\lambda))v(z')\mu(z')dx_1'dx_2'|
\]
\[
\leq C(\int_{\mathcal O}|g(z-z',k(\lambda))|^qdx_1'dx_2')^{1/q}\|v\|_{L^p}\sup|\mu|,
\]
where $p^{-1}+q^{-1}=1$.  Thus from (\ref{19DecD}) with an arbitrary $\alpha<2/q$ it follows that the integral on the right does not exceed $C_\alpha/(|\lambda|+1/| \lambda|)^\alpha$, i.e., the $L^\infty$-norm of the integral operator in (\ref{lsmu}) is less than one when $|\lambda|+1/ |\lambda|$ is sufficiently large. The latter implies the absence  of exceptional points.
\qed
\\

Since the solution $\psi(z,\lambda)$ of the Faddeev scattering problem may not exist for $\lambda\in D$, the following function $\psi'$ is considered.
For each $z \in \mathbb C$, function $\psi'$ is defined as follows:
\begin{equation}\label{pspr}
\psi' = \left \{ \begin{array}{l}
\psi(z,\lambda), \quad \lambda \in \mathbb C \backslash (D\bigcup\{0\}), \\
\psi^+(z,\lambda), \quad  \lambda \in D,
\end{array}
\right.
\end{equation}
where $\psi(z,\lambda)$ is the solution of the Faddeev scattering problem, and
function $\psi^+=\psi^+(z,\lambda)$ is the solution of the classical scattering problem  with the incident wave
$$
e^{ikx}=e^{\frac{i}{2} \sqrt{E} (\lambda \overline{z} + z /\lambda) },
\quad z \in \mathbb C,
$$
i.e., for each $\lambda \in \mathbb C\backslash\{0\}$, function $\psi^+$ satisfies the  Lippmann-Schwinger equation
\begin{equation}\label{eq1}
\psi^+ =e^{\frac{i}{2} \sqrt{E} (\lambda \overline{z} + z / \lambda) } +
\int_{\mathbb R^2}  G^+(z-z') v(z') \psi^+(z',\lambda)dx_1' dx_2',
\end{equation}
where
$$
G^+ =- \frac{1}{(2\pi)^2}  \int_{\xi \in \mathbb R^2} \frac{e^{i\xi {x}}}{|\xi|^2 - 2k \cdot \xi} d \xi = -\frac{i}{4}H^1_0(|z| \sqrt{E}).
$$
It is clear that $\psi^+$ is analytic in $\lambda \in \mathbb C\backslash \{0\}$.
We will also need the following function defined through $\psi'$:
\begin{equation}\label{mupr}
\mu'=\psi' e^{-i\frac{\sqrt{E}}{2} (\lambda \overline{z}+z / \lambda)}=\psi'(z,\lambda) \varphi_0(z,-\lambda).
\end{equation}

Next, we introduce an integral operator that will be used for reconstructing the potential. Let
 \begin{equation} \label{may1}
 e_0(z,\lambda)=e^{-i\frac{\sqrt{E}}{2} (\lambda \overline{z}+z / \lambda)}e^{-i\frac{\sqrt{E}}{2} (\frac{1}{\overline \lambda} \overline{z}+z \overline{\lambda})} = \varphi_0(z,-\lambda)\varphi_0(z,\frac{-1}{\overline{\lambda}}).
 \end{equation}
Consider the space
\[
\mathcal H^s :=  L^s(\mathbb R^2) \cap C(\overline{D}) , \quad s>2.
\]
Let $T_z:\mathcal H^s\to \mathcal H^s$ be the operator defined  by the formula
\begin{equation}\nonumber
T_z=T_z^{(1)} +T_z^{(2)} , \quad ~~ {\rm where} \quad T_z^{(1)} \phi=
-\frac{1}{\pi} \int_{\mathbb C }  r'(\varsigma) e_0(z,\varsigma) \phi(\frac{-1}{\overline \varsigma})   \frac{d\varsigma_R d\varsigma_I}{\varsigma-\lambda} ,
\end{equation}
\begin{equation}\label{19DecB}
T_z^{(2)} \phi=\frac{1}{2\pi i} \int_{\partial D} \frac{d\varsigma}{\varsigma - \lambda} \int_{\partial D} c(\varsigma,\varsigma')h(\varsigma',\varsigma) \varphi_0(z,-\varsigma) \varphi_0(z,\varsigma')\phi^-(\varsigma')d\varsigma', \quad \phi(\lambda)\in \mathcal H^s.
\end{equation}
Here $\phi^-$ is the  trace of $\phi$ on $\partial D$ taken from the interior of $D$, function $c(\varsigma,\varsigma')$ is given in~(\ref{4Jan1}), {\b functions $\varphi_0,e_0$ are defined in (\ref{exp}), (\ref{may1})}, and function $r'$ is defined by
\begin{equation}\label{Dec15C}
r'(\lambda)= \left \{ \begin{array}{l}
r(\lambda),  \quad \lambda \in \mathbb C \backslash (D\bigcup\{0\}), \\
0, \quad  \lambda \in D.
\end{array}
\right .
\end{equation}

The operator above can be used in the case of complex-valued potentials $v$. In the case of real-valued potentials, operator $T_z^{(1)}$ can be rewritten in a simpler form:
\begin{equation} \label{rreal}
 T_z^{(1)} \phi=
-\frac{1}{\pi} \int_{\mathbb C }  r'(\varsigma) e_0(z,\varsigma)\overline{\phi(\varsigma)}   \frac{d\varsigma_R d\varsigma_I}{\varsigma-\lambda}.
 \end{equation}

The following two theorems will be proved below.

\begin{theorem}\label{tmu}
Let a real-valued potential $v$ belong to $ L^{p}_{com}, p>1$. Then for each $z$, function (\ref{mupr}) satisfies the relation
\begin{equation} \nonumber
(I+T_z)(\mu'-1)=-T_z1, \quad \lambda\in \mathbb C \backslash \{0\}.
 \end{equation}
\end{theorem}

This theorem will allow us to prove the next statement.


Let us recall that a set $V$ of elements in a topological space $S$ is called {\it generic} if $V$ is open and dense in $S$.
\begin{theorem}\label{mthm}
Let the potential $v$ be real-valued and $v\in L^{p}_{com}, p>1$. Then
\begin{itemize}
\item
Operator $T_z$ considered in $\mathcal H^s, s>2$, is compact   for each $z \in \mathbb C$ and depends continuously on $z \in \mathbb C$.
\item
Function $T_z1$ belongs to $\mathcal H^s$ for each $ s >\widetilde{s}:=\max(q,4),~\frac{1}{p}+\frac{1}{q}=1,$ and depends continuously on $z$ (as element of $\mathcal H^s$).
  \item For every $z_0\in \mathbb C$ and generic  potentials $v\in L^{p}_{com}$, $p>1$, the equation
  \begin{equation}\nonumber
 (I+T_z)(\phi(z,\cdot)-1) =-T_z1, \quad \phi -1 \in \mathcal H^s, \quad s > \widetilde{s},
 \end{equation}
  is uniquely solvable for all $z$ in some neighborhood of $z_0$ in $\mathbb C$. The solution $\phi$ coincides with $\mu'$. Function $u=e^{ik{x}}\phi(z,\lambda)$ coincides with  $\psi'$ and satisfies the equation $(-\Delta-E+v(z))u=0$ in $\mathcal O$ for each $\lambda\in\mathbb C, ~|\lambda|\neq 0,1$.
 \item Potential $v$ can be found as $v=\frac{\Delta u}{u}+E$ or, if $v$ is smooth enough, calculated from the formula $v=-i\partial_{\overline{z}} a_1(z)$, where
\begin{eqnarray}\nonumber
a_1(z)=
-\frac{1}{\pi} \int_{\mathbb C }  r'\left ( \varsigma\right )e_0(z,\varsigma) \overline{\phi(z,\varsigma)}  d\varsigma_R d\varsigma_I \\ \nonumber +\label{6FevA}
\frac{1}{2\pi i} \int_{\partial D} d\varsigma \int_{\partial D} c(\varsigma,\varsigma')h(\varsigma,\varsigma')\varphi_0(z,-\varsigma) \varphi_0(\varsigma',z)\phi^-(z,\varsigma')d\varsigma'.
\end{eqnarray}
   \end{itemize}
\end{theorem}

{\bf Remark.} The generic set of potentials may depend on $z_0$, and the neighborhood of $z_0$ may depend on $v$.



\section{ Proof of Theorem \ref{tmu}}
\begin{lemma}\label{220616A}
Let $v \in L^{p}_{com},~ p>1$. Then $\widehat{g}:\mu \rightarrow g*(v\mu)$ is a compact operator in $L^\infty$. 
\end{lemma}
{\bf Proof.}  Faddeev's Green function $g$ has a logarithmic singularity, and therefore the convolution  $u:=g*(v\mu) $ belongs to $\in W^{2,p}_{loc}(\mathbb R^2)$.  Moreover,
\begin{equation}\label{xx11}
\|u\|_{ W^{2,p}_{loc}(\mathbb R^2)}\leq C(v)\sup|\mu|.
\end{equation}
Let us justify (\ref{xx11}) more rigorously. From (\ref{19DecD}) with small $\alpha>0$ it follows that $|u|\leq C(v)\sup|\mu|.$ Hence $\|u\|_{L^p_{loc}}\leq C(v)\sup|\mu|.$ Relation (\ref{2DecA}) between $G$ and $g$ implies that $g$ is a fundamental solution of the operator $\Delta+2ik\cdot \nabla$, and therefore
\[
(\Delta+2ik\cdot \nabla)u=v\mu, \quad x\in \mathbb R^2.
\]
From local elliptic a priory estimates it follows that for each $\rho>0$,
\[
\|u\|_{ W^{2,p}(|x|<\rho)}\leq C(\rho)(\|v\mu\|_{ L^{p}(|x|<\rho+1)}+\|u\|_{ L^{p}(|x|<\rho+1)}),
\]
and this immediately implies (\ref{xx11}).

Let $P_Rf=f$ when $|x|<R$, and  $P_Rf=0$ when $|x|>R$. Then (\ref{xx11}) and the Sobolev embedding theorem imply that for each $R$, the operator $P_R\widehat{g}$ is bounded as operator from $L^\infty$ to the Hölder's space $C^{2/q},~p^{-1}+q^{-1}=1$. Thus it is compact in $L^\infty$. From (\ref{19DecD}) with arbitrary $\alpha>0$ it follows that $|g*(v\mu)|\to 0$ as $|x|\to\infty$, and moreover $\|(I-P_R)\widehat{g}\|_{L^\infty}\to 0$ as $R\to\infty$. Thus $\widehat{g}$ is a limit of compact operators $P_R\widehat{g}$ as $R\to\infty$, and therefore it is compact.
\qed


It is not difficult to show that the function $\mathcal E=\frac{1}{\pi(\lambda-\lambda_0)}$ is a fundamental solution for the operator $\partial/\partial \overline{\lambda}$. i.e., $\frac {\partial}{\partial\overline{\lambda}}\mathcal E=\delta (\lambda-\lambda_0).$ This fact lies at the foundation of the following  important  result that will be used essentially in this paper (the proof can be found in the Appendix, part II, or \cite[(3.14)]{RN2}, \cite[lemma 3.1]{g2000}):
\begin{equation}\label{12NovD}
\frac{\partial}{\partial\overline{\lambda}} G(z,k(\lambda)) = \frac{\sgn (|\lambda|^2-1)}{4\pi \overline{\lambda}}
e^{-i\sqrt{E}/2 (\overline{\lambda} z + \overline{z} / \overline{\lambda})}, \quad |\lambda|\neq 0,1.
\end{equation}
 Since the  function $G(x,k)$ is real valued, the latter relation implies that
\begin{equation}\label{12NovDa}
\frac{\partial}{\partial\lambda} G(z,k(\lambda)) = \frac{\sgn (|\lambda|^2-1)}{4\pi \lambda}
e^{i\sqrt{E}/2 (\lambda \overline{z} + {z} / \lambda)} , \quad |\lambda|\neq 0,1.
\end{equation}

The following equation can be obtained (see e.g. \cite{gm}) by differentiating the Lippmann-Schwinger equation (\ref{19JanA}) in $\overline{\lambda}$ and using (\ref{12NovD}) (see details in the Appendix, part III):
\begin{equation}\label{dbar}
\frac{\partial}{\partial \overline{\lambda}} \psi (z,\lambda)=r(\lambda) \psi\left (z,-\frac{1}{\overline \lambda} \right ), \quad \lambda\notin \overline D\bigcup\{0\}.
\end{equation}
If the potential is real-valued, the latter equation can be replaced by a simpler one:
\begin{equation}\label{dbar1}
\frac{\partial}{\partial \overline{\lambda}} \psi (z,\lambda)=r(\lambda)\overline {\psi (z, \lambda  )}, \quad \lambda\notin \overline D\bigcup\{0\}.
\end{equation}
Note that the condition $\lambda\notin \overline D$ is essential here since $\psi$ is not smooth in $\lambda, \overline \lambda$ at the exceptional points for the equation (\ref{2DecA}). Thus we replace $\psi$ by function (\ref{eq1}) when $\lambda\in D$.
Since $\psi^+$ is analytic in $\lambda$, its derivative in $ \overline{\lambda}$ vanishes, and (\ref{dbar}) implies that
\begin{equation}\nonumber
\frac{\partial}{\partial \overline{\lambda}} \psi' (z,\lambda)=r'(\lambda) \psi'\left (z,\frac{-1}{\overline \lambda} \right ), \quad \lambda\notin \partial D\bigcup\{0\},
\end{equation}
where $\psi'$ is defined in (\ref{pspr}) and $r'$ is given by (\ref{Dec15C}).

This equation must be complemented by the boundary conditions on $\partial D$. The boundary conditions will be derived later. First we would like to express
function $\psi'$ via $\mu'$ using~(\ref{mupr}).

Using (\ref{mu1}) and (\ref{may1}), we can rewrite (\ref{dbar}) as follows:
\begin{equation}\nonumber
\frac{\partial}{\partial \overline{\lambda}} \mu (z,\lambda) =
r(\lambda)e_0(z,\lambda) \mu\left (z,-\frac{1}{\overline \lambda}\right ), \quad \lambda\notin \overline D\bigcup\{0\}.
\end{equation}
Since $\mu'$ is analytic in $D$, it follows that $\mu'$ satisfies the equation
\begin{equation}\label{mu4}
 \frac{\partial}{\partial \overline{\lambda}} \mu' (z,\lambda)=r'(\lambda)e_0(z,\lambda) \mu'\left (z,\frac{-1}{\overline \lambda}\right ), \quad \lambda\notin \partial D\bigcup\{0\}.
 \end{equation}

Equation (\ref{mu4}) is the main $\overline{\partial}$-equation that leads to the statement of Theorem \ref{tmu}. The advantage of considering $\mu'$ instead of $\psi'$ is due to the simple behavior of $\mu'$ at the origin and infinity (see \cite{gm}):
\begin{equation}\label{xx}
\lim_{\lambda \rightarrow \infty} \mu'(z,\lambda)=1, \quad \lim_{\lambda \rightarrow 0} \mu'(z,\lambda)=1
\end{equation}
uniformly in $|z|$. The latter is a consequence of the Lippmann-Schwinger equation (\ref{lsmu}) and~(\ref{19DecD}).
Equation (\ref{mu4}) can be rewritten in the following simpler form if the potential $v$ is real-valued:
\begin{equation}\label{mu4a}
 \frac{\partial}{\partial \overline{\lambda}} \mu' (z,\lambda)=r'(\lambda)e_0(z,\lambda) \overline{\mu'\left (z,\lambda \right )}, \quad \lambda\notin \partial D\bigcup\{0\}.
 \end{equation}
In order to justify this, one needs only to replace (\ref{dbar}) by (\ref{dbar1}) in the arguments above.

\begin{lemma}\label{lemma12NovA} The following relation holds for function $\mu'$ for each $z \in \mathbb C$ and $\lambda\notin \partial D\bigcup\{0\}$:
\begin{equation}\label{11NovC}
\mu'(z,\lambda)-1 = T_z^{(1)}\mu' + \frac{1}{2\pi i} \int_{\partial D} \frac{[\psi^+(z,\varsigma)-\psi(z,\varsigma)]\varphi_0(z,-\varsigma)}{\varsigma - \lambda}d\varsigma.
\end{equation}
\end{lemma}
{\bf Remark.} Here and throughout the paper, the direction of integration over the boundary of a domain is chosen in such a way that the domain remains on the left during the motion along the boundary.

{\bf Proof.} The following Cauchy-Pompeiu formulas hold for each $f\in C^1(\overline \Omega)$ and an arbitrary bounded domain $\Omega$ with a smooth boundary:
\begin{eqnarray}\label{2DecB}
f(\lambda)= - \frac{1}{\pi} \int_{ \Omega} \frac{\partial f (\varsigma)}{\partial \overline \varsigma} \frac{d\varsigma_R d\varsigma_I}{\varsigma-\lambda} + \frac{1}{2\pi i} \int_{\partial   \Omega} \frac{f(\varsigma)}{\varsigma - \lambda}d\varsigma, \quad \lambda \in  \Omega, \\ \label{2DecC}
0= - \frac{1}{\pi} \int_{ \Omega} \frac{\partial f (\varsigma)}{\partial \overline \varsigma} \frac{d\varsigma_R d\varsigma_I}{\varsigma-\lambda} + \frac{1}{2\pi i} \int_{\partial \Omega} \frac{f(\varsigma)}{\varsigma - \lambda}d\varsigma,\quad \lambda \not \in \overline{ \Omega}.
\end{eqnarray}

Let $D_R=\{\lambda\in \mathbb C: R^{-1}<|\lambda| < R\}$, i.e., $D_R$ is the ring $D$ with $A$ replaced by $R$. We will assume that $R>A$. Let $D_R^-=D_R\backslash D $. Assume that $\lambda\in D_R^-$. Then we take the sum of formulas (\ref{2DecB}) and (\ref{2DecC}) with $f=\mu'$ in both, and $\Omega=D_R^-$ in (\ref{2DecB}) and $\Omega=D$ in (\ref{2DecC}). If $\lambda\in D$, then we use (\ref{2DecB}) with $\Omega=D$ and (\ref{2DecC}) with $\Omega=D_R^-$. If we take (\ref{mu4}) and (\ref{mu4a}) into account, we obtain that
\begin{equation}\label{11NovA}
\mu'(\lambda) = T_z^{(1)}\mu' + \frac{1}{2\pi i} \int_{\partial D} \frac{[\mu']}{\varsigma - \lambda}d\varsigma+ \frac{1}{2\pi i} \int_{\partial D_R} \frac{\mu'}{\varsigma - \lambda}d\varsigma,
\end{equation}
where $[\mu']$ is the jump of $\mu'$ on $\partial D$, i.e., $[\mu']$ is the limiting value on $\partial D$ from the interior of $D$ minus the limiting value from the exterior of $D$. The statement of the lemma follows from (\ref{11NovA}) and (\ref{mupr}) if we take $R\to\infty$ and note that the last term on the right-hand side above converges to one due to (\ref{xx}).
\qed

Equation (\ref{11NovC}) does not take into account that the functions $\psi$ and $\psi^+$ are related. Our next goal is to take this relation into account and change the last term in (\ref{11NovC}). The first step in this direction is the following lemma.
\begin{lemma}\label{1DecLA}
Denote by $c=c(\lambda,\varsigma)$ the function
\[
c(\lambda,\varsigma) =  \frac{i}{2}{\rm sgn}(|\lambda |^2-1)[\frac{1}{\varsigma}{\rm Ln}\frac{\varsigma-\lambda}{\varsigma-\frac{\lambda}{|\lambda|}}+\varsigma{\rm Ln}\frac{\frac{-1}{\varsigma}-\overline{\lambda} } {\frac{-1}{\varsigma}-\frac{\overline{\lambda}}{|\lambda|}} ]
\]
\begin{equation}\label{4Jan1}
+\int_{|\varsigma_1|=1}\frac{1}{2(\varsigma-\varsigma_1)}  \theta \left  [i{\rm sgn}(|\lambda |^2-1) \left  (\frac{|\lambda|\varsigma_1}{\lambda}- \frac{\lambda}{|\lambda|\varsigma_1}\right )\right ]|d\varsigma_1|
, \quad \lambda,\varsigma \in \partial D,
\end{equation}
where $\theta$ is the Heaviside function. Then
\begin{equation}\label{13NovB}
G(z,k(\lambda))-G^+(z) =\frac{1}{(2\pi)^2} \int_{\partial D} c(\lambda,\varsigma, E) e^{i\sqrt{E}/2 (\varsigma\overline{z} + z / \varsigma)}d\varsigma, \quad~ \lambda,\varsigma \in \partial D.
\end{equation}
\end{lemma}
{\bf Remark.} For each $\lambda\in\overline {D}$, the segment $[\lambda,\frac{\lambda}{|\lambda|}]\subset\overline D $ is seen from each point $\varsigma\in\partial D$ under an angle $\psi$ such that $|\psi|<\pi$. Similarly, $[\overline\lambda,\frac{\lambda}{|\lambda|}]$ is seen from each point $\frac{-1}{\varsigma}$,
$\varsigma\in\partial D$, under an angle $\psi'$ such that $|\psi'|<\pi$. Thus
\begin{equation}\nonumber
|\arg f_i| <\pi,~~i=1,2,
\end{equation}
for the ratios $f_i$ under the logarithm signs in (\ref{4Jan1}), i.e., the logarithms are uniquely defined by the condition $|\Im {\rm Ln} f_i|<\pi$.

{\bf Proof.}
Using the Cauchy formula, one can rewrite (\ref{12NovD}),  (\ref{12NovDa}) in the form
$$
\frac{\partial}{\partial \overline{\lambda}} G(z,k(\lambda)) = \frac{-1}{2\pi i} \int_{\partial D} \frac{\sgn (|\lambda |^2-1)}{4\pi \overline{\varsigma}}
e^{-i\sqrt{E}/2 (\overline{\varsigma} {z} + \overline{z} / \overline{\varsigma})}\frac{d\overline{\varsigma}}{\overline{\varsigma}-\overline{\lambda}},  \quad \lambda\in \overline{D}, ~~|\lambda|\neq 1.
 $$
\begin{equation*}\label{2DecF1}
\frac{\partial}{\partial \lambda} G(z,k(\lambda)) = \frac{1}{2\pi i} \int_{\partial D} \frac{\sgn (|\lambda |^2-1)}{4\pi \varsigma}
e^{i\sqrt{E}/2 ({\varsigma} \overline{z} + {z} / \varsigma)}\frac{d\varsigma}{\varsigma-\lambda}, \quad  \lambda\in \overline{D},~~|\lambda|\neq 1.
 \end{equation*}
One can replace $G$ by $G-G^+$ here since $G^+$ does not depend on $\lambda, \overline{\lambda}$. One can reconstruct the function $G-G^+$ in the domain $|\lambda|>1~(|\lambda|<1)$ from the potential $u_+~(u_-$ respectively) of its gradient field with respect to variables $\overline{\lambda}, \lambda$:
\begin{equation}\label{rec}
G-G^+=u_\pm(z,\lambda)-u_\pm(z,\lambda_0^\pm)+[G(z,k(\lambda_0^\pm))-G^+(z)],  \quad \lambda\in D, ~~|\lambda|\gtrless 1,
 \end{equation}
where $\lambda_0^\pm$ is an arbitrary point in the domain $|\lambda|\gtrless 1$ where the gradient field is defined.

We choose $\lambda_0^\pm=\frac{\lambda}{|\lambda|}(1\pm 0)$ since the limiting values of  $G-G^+$ on the unit circle $|\lambda|=1$ are found in \cite[section 3]{RN2}:
$$
 G(z,k(\lambda_0^\pm))-G^+(z) = \frac{\pi i}{(2\pi)^2}  \int_{|\varsigma|=1} e^{i\sqrt{E}/2 (\varsigma\overline{z} + z / \varsigma)} \theta \left  [ i \sgn (|\lambda |^2-1)\left   (\frac{|\lambda|\varsigma}{\lambda}- \frac{\lambda}{|\lambda|\varsigma}\right )\right ]|d \varsigma|,
 $$
 where $\theta$ is the Heaviside function. Using the Cauchy formula, we can rewrite the latter equality as follows:
 $$
G(z,k(\lambda_0^\pm))-G^+(z)
$$
\begin{equation}\label{121}
=\frac{1}{8\pi^2}\int_{|\varsigma_1|=1} \left( \int_{\partial D} \frac{ e^{i\sqrt{E}/2 (\varsigma\overline{z} + z / \varsigma)} d\varsigma}{\varsigma-\varsigma_1} \right ) \theta \left  [ i \sgn (|\lambda |^2-1)\left  (\frac{|\lambda|\varsigma_1}{\lambda}- \frac{\lambda}{|\lambda|\varsigma_1}\right )\right ]|d\varsigma_1|.
\end{equation}

One can easily check that functions
\[
u_\pm=-\frac{\sgn (|\lambda |^2-1)}{8\pi^2 i} \int_{\partial D} \frac{1}{ \overline{\varsigma}}
e^{-i\sqrt{E}/2 (\overline{\varsigma} {z} + \overline{z} / \overline{\varsigma})}{\rm Ln}(\overline{\varsigma}-\overline{\lambda})d\overline{\varsigma}
\]
\begin{equation}\label{upm}
-\frac{\sgn (|\lambda |^2-1)}{8\pi^2 i} \int_{\partial D} \frac{1}{ \varsigma}
e^{i\sqrt{E}/2 ({\varsigma} \overline{z} + {z} / \varsigma)}{\rm Ln}(\varsigma-\lambda)d\varsigma, \quad \lambda\in \overline{D},~~|\lambda|\gtrless 1,
\end{equation}
are potentials of the field (\ref{12NovD}), (\ref{12NovDa}). The logarithms here are defined as follows. We fix a negative $\lambda\in D$ and positive values of $\varsigma=A, ~1/A$ on the connected components of $\partial D$, and choose the logarithms to be real-valued at these points. The values at all other points $(\varsigma, \lambda),~\varsigma\in \partial D, \lambda \in \overline{D},$ are obtained by analytic continuation. We can also impose the condition $\arg |\varsigma|<\pi$ in order to avoid a discussion about possible branching of the logarithms when $\varsigma$ goes along either of the circles that are components of $\partial D$. The logarithms remain multi-valued functions of $\lambda$: their values change by $\pm2\pi i$ when $\lambda$ travels along a closed simple curve in $\overline{D}$ around the origin. However, $u_\pm$ are well defined single-valued functions since the integrals in (\ref{upm}) with the logarithms replaced by a constant are equal to zero due to the analyticity in $\overline{D}$ of the integrands in (\ref{upm}) when
the logarithms are replaced by constants.

 We have
 \[
u_\pm(z,\lambda)-u_\pm(z,\lambda_0^\pm)=-\frac{\sgn (|\lambda |^2-1)}{8\pi^2 i} \int_{\partial D} \frac{1}{ \overline{\varsigma}}
e^{-i\sqrt{E}/2 (\overline{\varsigma} {z} + \overline{z} / \overline{\varsigma})}{\rm Ln}\frac{\overline{\varsigma}-\overline{\lambda}}{\overline{\varsigma}-\frac{\overline{\lambda}}{|\lambda|}}d\overline{\varsigma}
\]
\[
-\frac{\sgn (|\lambda |^2-1)}{8\pi^2 i} \int_{\partial D} \frac{1}{ \varsigma}
e^{i\sqrt{E}/2 ({\varsigma} \overline{z} + {z} / \varsigma)}{\rm Ln}\frac{\varsigma-\lambda}{\varsigma-\frac{\lambda}{|\lambda|}}d\varsigma, \quad \lambda\in \overline{D},~~|\lambda|\gtrless 1.
\]

Due to the remark after the lemma, we can forget now about possible branching of the logarithms and define the values of the logarithms by the condition $|\Im {\rm Ln}(\cdot)|<\pi/2$.
 We change the variable $\varsigma\to\frac{-1}{\overline{\varsigma}}$ in the first integral on the right and put the resulting formula and (\ref{121}) into (\ref{rec}). This proves the statement of the lemma.
\qed

Now we can express $\psi-\psi^+$ in (\ref{11NovC}) as an image of a compact operator applied to $\psi_+$.
\begin{lemma}\label{13NovA}
The following representation holds
\begin{equation}\nonumber
\psi(z,\lambda)= \psi^+(z,\lambda) + \int_{\partial D} c(\lambda,\varsigma)h(\varsigma,\lambda)\psi^+(z,\varsigma)d\varsigma, \quad \lambda \in \partial D.
\end{equation}
where $c(\lambda,\varsigma)$ is given by formula (\ref{4Jan1}) and $h$ is defined in (\ref{27dec2}).
\end{lemma}
{\bf Proof.}
Recall that $\varphi_0(z,\lambda) = e^{i/2 \sqrt{E} (\lambda \overline{z} + z/ \lambda)}$. We will use notation $G^+, G$ not only for the Green functions, but also for for the convolution operators with the kernels $G^+, G$. We will denote by $G^+v, Gv$ the operator of multiplication by the potential $v$ followed by the convolution $G^+$ or $G$, respectively. Then one can rewrite (\ref{eq1}) and (\ref{19JanA}) as follows:
\begin{equation}\label{prp1}
\psi^+(z,\lambda)= (I-G^+v)^{-1} \varphi_0, \quad
\psi(z,\lambda)= (I-Gv)^{-1} \varphi_0.
\end{equation}
Thus
$$
\psi^+(z,\lambda)= (I-G^+v)^{-1} [(I-G v)\psi(z,\lambda)],
$$
and therefore
\begin{equation}\label{prp}
\psi(z,\lambda)-\psi^+(z,\lambda) = (I-G^+v)^{-1} (G - G^+)(v(\cdot) \psi(\cdot,\lambda) ).
\end{equation}

We express $G - G^+$ via (\ref{13NovB}) and use the relation $\varphi_0(z-u,\lambda)=\varphi_0(z,\lambda)\varphi_0(-u,\lambda)$. This leads to
\begin{eqnarray*}
(G - G^+)(v(\cdot) \psi(\cdot,\lambda) )\\ = \frac{1}{(2\pi)^2}  \int_{\partial D}  \int_{\mathbb R^2}c(\lambda,\varsigma) \varphi_0(z,\varsigma)\varphi_0(-u,\varsigma)d\varsigma v(u) \psi(u,\lambda)du_Idu_R\\=  \int_{\partial D} c(\lambda,\varsigma) \varphi_0(z,\varsigma)h(\varsigma,\lambda)d\varsigma .
\end{eqnarray*}
We plug the last relation into (\ref{prp}). It remains to note (see (\ref{prp1})) that $(I-G^+v)^{-1}\varphi_0(\cdot,\varsigma) =\psi^+(z,\varsigma).$
\qed
\\

Theorem \ref{tmu} is a direct consequence of Lemmas \ref{lemma12NovA} and \ref{13NovA}.
\section{Proof of Theorem \ref{mthm}.}
From now on we will consider only real-valued potentials since  more complicated arguments and an additional smoothness assumption are needed to treat complex-valued potentials. Let us prove the statement on the compactness of the operator
\[
T_z=T_z^{(1)}+T_z^{(2)}:\mathcal H^s\to \mathcal H^s, \quad  \mathcal H^s= L^s(\mathbb C) \cap C(\overline{D}) , \quad s>2,
\]
where $T_z^{(1)}, T_z^{(2)}$ are defined in (\ref{19DecB}), (\ref{rreal}). We will write operator (\ref{rreal}) in the form
 \begin{equation}\label{dmin}
  T_z^{(1)}f=\overline{\partial}^{\!~-1}(r'e_0\overline{f}), \quad {\rm where} \quad \overline{\partial}^{\!~-1}\phi = -\frac{1}{\pi} \int_{\mathbb C }  \phi(\varsigma)   \frac{d\varsigma_R d\varsigma_I}{\varsigma-\lambda}.
  \end{equation}
  where $r'$ is defined in (\ref{Dec15C}).

The following lemma provides an estimate on the factor
$$
t(z,\varsigma)= r' (\varsigma) e_0(z,\varsigma)
$$
in the integral kernel of the operator $T_z^{(1)}$, see (\ref{dmin}), (\ref{Dec15C}).

Let $p>1$ and $s_1(p)=\frac{2\widetilde{s}}{\widetilde{s}+2},~s_2(p)=\frac{2\widetilde{s}}{\widetilde{s}-2},$ where $\widetilde{s}=\max(q,4),~p^{-1}+q^{-1}=1 $. Obviously,
\[
1<s_1(p)<2<s_2(p).
\]
\begin{lemma}\label{llp2}
If $v\in L^p_{com},~p>1$, then
$t(z,\cdot) \in L^{s}(\mathbb C)$ for each $s$ in the interval $s_1(p)<s<s_2(p)$. Moreover, $\|t(z,\cdot)\|_{L^{s}(\mathbb C)}$ does not depend on $z$, and
function $t$, as element of $L^{s}(\mathbb C)$, depends continuously on $z\in \mathbb C$.
\end{lemma}
{\bf Proof.}
Let us prove that $ r'(\cdot) \in L^{s}(\mathbb C\backslash D)$. In order to obtain this inclusion, we express $\psi$ in (\ref{27dec1}) via $\mu$ using (\ref{mu1}) and then split $r(\lambda)$ in formula (\ref{27dec1}) into two terms by writing $\mu$ in the integrand as $\mu=1+(\mu-1)$. This and (\ref{Dec15C}) lead to
$$
\frac{4\pi r'(\lambda)}{\sgn(|\lambda|^2-1)}= \frac{1}{\lambda}\int_{\mathbb R^2} e^{ i\frac{\sqrt{E}}{2} \Re  [ z  (\overline{\lambda} + \frac{1}{\lambda} )]} v(z) dx_1dx_2+ \frac{1}{\lambda}\int_{\mathbb R^2} e^{ i\frac{\sqrt{E}}{2} \Re  [ z  (\overline{\lambda} + \frac{1}{\lambda} )]} v(z) (\mu(z,\lambda)-1) dx_1dx_2
$$
\begin{equation}\label{1234}
=r_1+r_2, \quad  \lambda\notin \overline{D}\bigcup\{0\}.
\end{equation}
We put $r_1=r_2=0$ when $\lambda\in D$ (recall that $r'(\lambda)=0$ in $D$).

Let us prove that $r_1(\cdot) \in L^{s}(\mathbb C )$. Note that
$$
r_1(\lambda)=\frac{1}{\lambda}\widehat{v}(\omega ), \quad \omega=\frac{\sqrt{E}}{2}(\lambda + \frac{1}{\overline{\lambda}} ),
$$
where $\widehat{v}$ is the Fourier transform, with the real and imaginary parts of $-\omega$  being the dual variables to $(x_1,x_2)$. Thus the inclusion $ r_1(\cdot) \in L^{s}(\mathbb C )$ is equivalent to
\begin{equation}\label{incl}
\int_{|\lambda|>1} \frac{|\widehat{v}(\omega )|^{s}}{|\lambda|^{s}} d\lambda_\Re d \lambda_\Im+ \int_{|\lambda|<1} \frac{|\widehat{v}(\omega )|^{s}}{|\lambda|^{s}} d\lambda_\Re d \lambda_\Im:=a_1+a_2<\infty.
\end{equation}

The proof of (\ref{incl}) will be based on the Hausdorff-Young inequality stating that $\widehat{v}=\widehat{v}(\omega)\in L^q,~p^{-1}+q^{-1}=1$ if $1<p\leq2$. The Jacobian of the map $\omega=\frac{\sqrt{E}}{2}(\lambda + \frac{1}{\overline{\lambda}} )$ in $\mathbb C$ is equal to $\frac{E}{4}|1-\frac{1}{|\lambda|^4}|$. Hence
\[
\int_{\mathbb C} |1-\frac{1}{|\lambda|^4}||\widehat{v}(\omega )|^{q} d\lambda_\Re d \lambda_\Im <\infty.
\]
Since $v$ is compactly supported, function $\widehat{v}$ is smooth in $\omega$. Thus the factor $|1-\frac{1}{|\lambda|^4}|$ above can be replaced by $1+\frac{1}{|\lambda|^4}$, i.e.,
\begin{equation}\label{vbn}
\int_{\mathbb C} |1+\frac{1}{|\lambda|^4}||\widehat{v}(\omega )|^{q} d\lambda_\Re d \lambda_\Im <\infty, \quad 1<p\leq 2.
\end{equation}
If $v\in L^{p}_{com}$ with some $p>1$, then $v$ belongs to the same space with any smaller value of $p>1$. In particular, if $p>4/3$, then $v\in L^{4/3}_{com}$, and therefore (\ref{vbn}) holds with $q=4$ (which is dual to $p=4/3$). In other words, (\ref{vbn}) holds for each $p>1$ if $q$ is replaced by $\widetilde{s}=\max(q,4)$.  Hence for each $p>1$,
\begin{equation}\label{vbn2}
\int_{|\lambda|>1}|\widehat{v}(\omega )|^{\widetilde{s}} d\lambda_\Re d \lambda_\Im +\int_{|\lambda|<1}\frac{1}{|\lambda|^4}|\widehat{v}(\omega )|^{\widetilde{s}} d\lambda_\Re d \lambda_\Im :=b_1+b_2<\infty.
\end{equation}
The terms $a_1,a_2$ in the left-hand side of (\ref{incl}) can be estimated by the corresponding terms $b_1,b_2$ in the left-hand side of (\ref{vbn2}) using the H$\ddot{o}$lder inequality. Indeed, for $s<\widetilde{s}$, we have\[
a_1=\int_{|\lambda|>1}|\widehat{v}(\omega )|^{s}|\lambda|^{-s} d\lambda_\Re d \lambda_\Im  \leq b_1^{\frac{s}{\widetilde{s}}}(\int_{|\lambda|>1} |\lambda|^{-\tau} d\lambda_\Re d \lambda_\Im)^{\frac{\widetilde{s}-s}{\widetilde{s}}}<\infty, \quad \tau=\frac{ s\widetilde{s}}{\widetilde{s}-s},
\]
since  $s<s_2(p)$ implies that $s<\widetilde{s}$, and the latter inequality together with $s>s_1(p)$ imply that $\tau>2$.

The term $a_2$ can be estimated similarly:
\[
a_2=\int_{|\lambda|<1} \frac{|\widehat{v}(\omega )|^{s}}{|\lambda|^{4 s/\widetilde{s}}}|\lambda|^{\frac{4s}{\widetilde{s}}-s} d\lambda_\Re d \lambda_\Im  \leq b_2^{\frac{s}{\widetilde{s}}}(\int_{|\lambda|<1} |\lambda|^{\tau} d\lambda_\Re d \lambda_\Im)^{\frac{\widetilde{s}-s}{\widetilde{s}}}<\infty, \quad \tau=\frac{4s -s\widetilde{s}}{\widetilde{s}-s},
\]
since $s<s_2(p)$ implies that $s<\widetilde{s}$, and these two inequalities together lead to $\tau>-2$. Thus $r_1(\cdot) \in L^{s}(\mathbb C \backslash D)$.

Let us prove that $r_2(\cdot) \in L^{s}(\mathbb C\backslash D)$.
It was shown in the proof of Lemma \ref{l21} that for each $\alpha\in (0,1],$ the norm in the space $L^\infty$ of the integral operator $\widehat g $ in the right-hand side of (\ref{lsmu}) does not exceed $C_\alpha(|\lambda|+|1\backslash \lambda|)^{-\alpha}$. Hence from (\ref{lsmu}) it follows that
\begin{equation}\label{eq10}
|\mu-1|\leq C_\alpha(|\lambda|+|1\backslash \lambda|)^{-\alpha}, \quad   |\lambda|+|1\backslash \lambda|\gg1, \quad 0<\alpha\leq 1.
\end{equation}
Recall that all the exceptional points belong to $D$, i.e., equation (\ref{lsmu}) is uniquely solvable in $L^\infty$. Since operator $\widehat g $ depends continuously on $\lambda\in \mathbb C\backslash D$, from the solvability of (\ref{lsmu}) for  $\lambda\in \mathbb C\backslash (D\bigcup\{0\})$ it follows that the solution depends continuously on $\lambda\in \mathbb C\backslash (D\bigcup\{0\})$, i.e., $|\mu|$ is uniformly in $\lambda$ bounded when $|\lambda|+|1\backslash \lambda|$ is bounded. Hence  (\ref{eq10}) holds for all $\lambda\in \mathbb C\backslash D$.
Since the support of $v$ is bounded, (\ref{1234}), (\ref{eq10}) imply that
\[
|r_2|(\lambda)\leq \frac{C_\alpha}{|\lambda|(|\lambda|+|1/\lambda|)^\alpha}, ~~~\lambda\in \mathbb C.
\]
 This estimate with $\alpha=1$ (or small enough $1-\alpha>0$) immediately implies that $ r_2(\cdot) \in L^{s}(\mathbb C\backslash D)$. Hence
\begin{equation}\label{rrrr}
r'(\cdot) \in L^{s}(\mathbb C), ~s_1(p)<s<s_2(p).
\end{equation}

Since $|e_0|=1$, the last inequality implies that $\|t(z,\cdot)\|_{L^{s}(\mathbb C)}$ does not depend on $z$. The arguments used above to prove (\ref{rrrr}) can be repeated to show that the inclusion (\ref{rrrr}) is valid for function $r'(\cdot)f_\varepsilon(\cdot)$, where $f_\varepsilon(\lambda)=(|\lambda|+\frac{1}{|\lambda|})^\varepsilon$ and $\varepsilon>0$ is small enough. Hence $t(z,\cdot)$ is a product of $z$-independent function $r'(\cdot)f_\varepsilon(\cdot)\in L^{s}(\mathbb C)$ and function $f_\varepsilon^{-1}(\cdot)e_0(z,\cdot)$, which is continuous in $z$ uniformly in $\lambda$. Hence  $t(z,\cdot)$ is continuous in $z$ as element of $L^{s}(\mathbb C)$.

 The proof of Lemma
\ref{llp2} is complete.
 \qed
\\

Let us complete the proof of Theorem \ref{tmu}.
For each function $g(\cdot)$ in $L^2(\mathbb C)$, the operator  $f\to\overline{\partial}^{\!~-1} (gf)$ is compact on $L^s(\mathbb C)$ for each $s>2$, and its norm does not exceed $C\|g\|_{L^2(\mathbb C)}$ (see, e.g., \cite[Lemma 3.1]{perry}, \cite[Lemma 5.3]{music} or Appendix, part IV).
From this fact and  Lemma \ref{llp2}, it follows that the operator $T_z^{(1)}$ is compact in $L^s(\mathbb C), s>2,$ and depends continuously on $z$. Thus the compactness of operator $T_z^{(1)}$ in $\mathcal H^s$ will be proved if we show its compactness as an operator from $L^s(\mathbb C)$ to $C(\overline{D})$.

We represent operator $T_z^{(1)}$ as the sum $P_\varepsilon+Q_\varepsilon$, where the terms are defined as follows:
\begin{equation}\label{pqpq}
P_\varepsilon f= -\frac{1}{\pi} \int_{D^\varepsilon\backslash D}  r'(\varsigma)e_0(z,\varsigma)\overline{f}(\varsigma)   \frac{d\varsigma_R d\varsigma_I}{\varsigma-\lambda}, \quad  Q_\varepsilon f=-\frac{1}{\pi} \int_{\mathbb C\backslash D^\varepsilon }  r'e_0\overline{f}   \frac{d\varsigma_R d\varsigma_I}{\varsigma-\lambda}.
\end{equation}
Here $D^\varepsilon$ is the $\varepsilon$-extension of the domain $D$, and the integration over $D$ is not involved in formulas above since $r'=0$ in $D$. Function $r'$ is smooth in $\mathbb C \backslash (D\bigcup\{0\})$, and therefore it is bounded in $D^\varepsilon\backslash D,~\varepsilon<1$. Thus
\[
|P_\varepsilon f|\leq C \|f\|_{L^s(\mathbb C)}(\int_{D^\varepsilon\backslash D}\frac{d\varsigma_R d\varsigma_I}{|\varsigma-\lambda|^{s'}})^{1/s'}, \quad \frac {1}{s}+\frac{1}{s'}=1, \quad \lambda\in \mathbb C.
\]
Since $s>2$ and the domain $D^\varepsilon\backslash D$ is shrinking as $\varepsilon \rightarrow 0$, it follows that $|P_\varepsilon f|\leq \alpha(\varepsilon) \|f\|_{L^s(\mathbb C)}$, where $\alpha(\varepsilon)$ vanishes as $\varepsilon\to 0$.

Let $\lambda\in D^{\varepsilon/2}$. Then from the inclusion $r'e_0\in L^2(\mathbb C)$ (see Lemma \ref{llp2}) and the Holder inequality  it follows that
\[
|Q_\varepsilon f|\leq C\|r'e_0\|_{L^2(\mathbb C)}
(\int_{\mathbb C\backslash D^\varepsilon } |f|^2(\varsigma)  \frac{d\varsigma_R d\varsigma_I}{|\varsigma-\lambda|^2})^{1/2}\leq C \|f\|_{L^s(\mathbb C)}(\int_{\mathbb C\backslash D^\varepsilon }\frac{d\varsigma_R d\varsigma_I}{|\varsigma-\lambda|^{2s/(s-2)}})^{\frac{s-2}{2s}}
\]
\[
\leq C(\varepsilon)\|f\|_{L^s(\mathbb C)}, \quad \lambda\in D^{\varepsilon/2}.
\]
It is also obvious that functions $Q_\varepsilon f$ are analytic in $\lambda\in D^{\varepsilon}$.

Since the uniform boundedness and analyticity of a set of functions in a bounded domain of the complex plane imply the pre-compactness of the set in the space $C$, it follows that the operator $Q_\varepsilon:L^s(\mathbb C)\to C(\overline{D})$ is compact. Hence $T_z^{(1)}:L^s(\mathbb C)\to C(\overline{D})$ is the limit as $\varepsilon\to 0$ of compact operators $Q_\varepsilon$, and therefore is compact. Its continuity in $z$ follows from the fact that $r'e_0$ is continuous in $z$ in the space $L^2(\mathbb R^2)$ (see Lemma \ref{llp2}).

Let us show the compactness and the continuity in $z$ of the second term  $T_z^{(2)}$ in the right-hand side of (\ref{19DecB}). We write $T_z^{(2)}$ in the form $T_z^{(2)}=I_1I_2R$, where $R:\mathcal H^s\to C(\partial D)$ is a bounded operator that maps a function $\phi\in \mathcal H^s$ into its boundary trace $\phi^-$ on $\partial D$ from the interior of $D$ (recall that $\phi$ belongs to $C(\overline{D})$), $I_2: C(\partial D) \rightarrow C^{\alpha}(\partial D)$ is the integral operator corresponding to the interior integral in the expression for $T_z^{(2)}$, and operator $I_1:C^{\alpha}(\partial D)\to \mathcal H^s$ is the integral operator corresponding to the exterior integral in the expression for $T_z^{(2)}$ (including the factor ${1}/{2\pi i}$).  Here $C^{\alpha}(\partial D)$ is the Holder space and  $\alpha$ is an arbitrary number in $(0,1/2)$.  The integral kernel of $I_2$ has a logarithmic singularity at $\varsigma=\varsigma'$ (due to the presence of the term $c(\varsigma,\varsigma')$). Thus operator $I_2$ is a PDO of order $-1$, and therefore $I_2$ is a bounded operator from $C(\partial D)$ into the  Sobolev space $H^1(\partial D)$. Hence it is compact as an operator from $C(\partial D)$ to $C^\alpha(\partial D), \alpha \in (0,1/2)$, due to the Sobolev imbedding theorem. Thus the compactness of  $T_z^{(2)}$ will be proved as soon as we show that $I_1$ is bounded.

For each $\phi \in C^\alpha(\partial D)$, function $I_1\phi$ is analytic outside of $\partial D$ and vanishes at infinity. Due to the Sokhotski-Plemelj theorem, the limiting values $(I_1\phi)_\pm$ of $(I_1\phi)$ on $\partial D$ from inside and outside of $D$, respectively, are equal to $\frac{\pm\phi}{2}+P.V.\frac{1}{2\pi i} \int_{\partial D} \frac{\phi(\varsigma)d\varsigma}{\varsigma - \lambda}$. Thus
\[
\max_{\partial D}|(I_1\phi)_\pm|\leq C\|\phi\|_{C^\alpha(\partial D)}.
\]
From the maximum principle for analytic functions, it follows that the same estimate is valid for the function $I_1\phi$ on the whole plane. Taking also into account that $I_1\phi$ has order $1/\lambda$ at infinity, we obtain that $|I_1 \phi| \leq \frac{C}{1+|\lambda|}\|\phi\|_{C^\alpha(\partial D)}$, i.e.,  operator $I_1$ is bounded. Hence operator $T_z^{(2)}$ is compact.

Obviously, operator $I_2$ depends continuously on $z$, and operators $R$ and $I_1$ do not depend on $z$, i.e., $T_z^{(2)}$ is continuous in $z$. The first statement of the theorem is proved.

Let us prove the second statement.
From the Hardy-Littlewood-Sobolev inequality it follows that
\[
\|T^{(1)}_z1\|_{L^{s}(\mathbb C)}\leq C\|r'\|_{L^{q}(\mathbb C)}, \quad \frac{1}{q}-\frac{1}{s}=\frac{1}{2}.
\]
Lemma \ref{llp2} allows us to choose an arbitrary $q$ from the interval $(s_1(p),2)$, and therefore $T^{(1)}_z1\in L^{s}(\mathbb C)$ with $s=\frac{2q}{2-q}$. Hence $s$ is an arbitrary number such that $s>\frac{2s_1(p)}{2-s_1(p)}=\widetilde{s}$. From Lemma \ref{llp2}, (\ref{pqpq}),  and the boundedness of $r'$ in every bounded region, it follows that
\[
|P_\varepsilon 1|\leq C\int_{D^\varepsilon\backslash D}\frac{d\varsigma_R d\varsigma_I}{|\varsigma-\lambda|}\leq C,  \quad |Q_\varepsilon 1|\leq C
\|r'e_0\|_{L^\alpha(\mathbb C)}
(\int_{\mathbb C\backslash D^\varepsilon }   \frac{d\varsigma_R d\varsigma_I}{|\varsigma-\lambda|^\beta})^{1/\beta}\leq C, \quad \lambda\in D,
\]
where $s_1(p)<\alpha<2,~\frac {1}{\alpha}+\frac {1}{\beta}=1$. The same arguments can be applied to show that $P_\varepsilon$ and $Q_\varepsilon$ are continuous in $\lambda\in \overline{D}$. Hence $T^{(1)}_z1\in C(\overline{D})$, and therefore $T^{(1)}_z1\in \mathcal H^{s}(\mathbb C)$ for each $s>\widetilde{s}$.

Now let us show that $T_z^{(2)}1 \in \mathcal H^{s}(\mathbb C)$ for each $s>\widetilde{s}$.
We have $T_z^{(2)}1=I_1I_2R1$, where operators $I_1,I_2,R$ were introduced earlier in the proof of the compactness. Since $R1 \in C(\partial D)$, the inclusion follows from the boundedness of operators $I_1,I_2$.

The continuity of $T_z1 \in \mathcal H^{s}(\mathbb C),~s>\widetilde{s},$ in $z$ follows from the continuity of $t(z,\cdot)$ as element of $L^s(\mathcal C)$, $s_1<s<s_2,$ (Lemma \ref{llp2}) and the uniform continuity of $t(z,\lambda)$ in each disk $|\lambda|<R$. The proof is the same as the proof of the continuity of operator $T_z$  in the space $\mathcal H^{s}(\mathbb C),~s>2$.

Let us prove the third statement of the theorem. We fix a point $z' \in \mathbb C$. The invertibility of $I+T_z$ at $z=z'$ implies its invertibility for $|z-z'|\ll 1$. Thus it is enough to show that the set $V\subset L^{p}(\mathcal O)$, $p>1$, of potentials $v$ for which $I+T_{z'}$ is invertible, is generic, i.e., this set is open and everywhere dense in the topology of $L^{p}(\mathcal O), p>1$. Obviously, operator $T_{z'}$ depends continuously on $v$. This can be proved by the same arguments that were used to prove the compactness of $T_z$. Hence if  $I+T_{z'}$ is invertible, then the same is true for a slightly perturbed potential, i.e., the set $V$ is open.

 If the invertibility is violated for a potential $v$, consider the set of potentials $av, a\in R$. Operator $T_{z'}$ is analytic in $a$ (see details in \cite[Section 4.2]{NLV} and \cite[Section 5]{lv15} if needed), and  $I+T_{z'}$ is invertible for small $a$, see \cite{gm}. Thus invertibility can be violated only in a set of isolated values of $a$. In particular, $I+T_{z'}$ is invertible for $0<|1-a|\ll 1$, i.e., the set $V$ is dense. Hence operator $I+T_{z'}$ is invertible  for a generic set of potentials.
The remaining part of the third statement of Theorem \ref{mthm} follows immediately from Theorem~\ref{tmu}.

 The proof of the last statement of the theorem is absolutely similar to proof of formula (36) in \cite{g2000}.

\qed

\section{Appendix}

{\bf I. Proof of (\ref{dddd}).} It is convenient to study the integrand in (\ref{2DecA}) using the complex variable $\eta=\xi_1+i\xi_2\in\mathbb C$ instead of $\xi \in \mathbb R^2$. Denote
\begin{equation}\label{lala}
P:=|\xi|^2+2k(\lambda)\cdot \xi=|\eta|^2 + \sqrt E(\lambda \overline{\eta}+ \frac{\eta}{\lambda}), \quad |\lambda|\neq0,1.
\end{equation}

Let us find all the points $\eta=\eta(\lambda)$ where $P=0$. Since $\lambda\neq 0$, we can make a substitution $\eta(\lambda)=\lambda c(\lambda)$. This leads to the following equation for the unknown $c$:
\begin{equation}\label{cca}
|c|^2|\lambda|^2+\sqrt E(\overline c|\lambda|^2+c)=0.
\end{equation}
Equating the imaginary part of the left-hand side to zero, we obtain that $\Im c(-|\lambda|^2+1)=0$. Hence $\Im c=0$ since $|\lambda|\neq1$ in (\ref{lala}). Now equation (\ref{cca}) becomes a simple quadratic equation for $c$ with the roots $0$ and $-\sqrt E-\frac{\sqrt E}{|\lambda|^2}$, i.e., $P=0$ at two points: $\eta=0$ and $\eta=\eta_0(\lambda)=-\sqrt E(\lambda+\frac{1}{\overline{\lambda}})$.

Let $\Lambda=\{\lambda:~|\lambda|^2+\frac{1}{|\lambda|^2}=4\}$. The two estimates below are valid for an arbitrary $E=E_0>0$ and an arbitrary compact set in the complex $\lambda$-plane that does not contain $\lambda=0$ and points with $|\lambda|=1$. However, it is sufficient for us to prove these estimates when $\lambda\in\Lambda, E=1/2.$

Let us show that there are positive constants $\gamma,\rho$ such that
\begin{equation}\label{awas}
|P|\geq\gamma|\eta|  \quad {\rm when} \quad \lambda\in \Lambda, ~~|\eta|\leq\rho,~~E=1/2.
\end{equation}
Indeed, the linear in $\overline{\eta},\eta$ part of $P$ is equal to $P_1=\lambda \overline{\eta}+ \frac{\eta}{\lambda}$. Note that a function $f:=a\overline{\eta}+b\eta=0$ with non zero complex constants $a$ and $b$ vanishes only at the origin if $|a|\neq |b| $ (since $f=0, \eta\neq 0,$ implies that $|\frac{b}{a}|=|\frac{\overline{\eta}}{\eta}|=1$). Hence $P_1\neq 0$ when $\eta\neq 0$ and $|\lambda|\neq0,1$. Now from the homogeneity of $P_1$  it follows that there is a constant $\gamma$ such that $|P_1|\geq 2\gamma|\eta|$ when $\lambda\in \Lambda,~\eta\in \mathbb C,~E=1/2$. This implies (\ref{awas}) if $\rho$ is small enough.

The same argument can be used to prove a similar estimate in a neighborhood of $\eta_0(\lambda)$:
\begin{equation}\label{awas1}
|P|\geq\gamma|\eta-\eta_0(\lambda)|  \quad {\rm when} \quad \lambda\in \Lambda, ~~|\eta-\eta_0(\lambda)|\leq\rho,~~E=1/2.
\end{equation}
In order to justify (\ref{awas1}), we need only to show that $|a|\neq|b|$ for $(a,b):=\nabla_{\overline{\eta},\eta}P|_{\eta=\eta_0(\lambda)}$.  By evaluating the gradient $\nabla_{\overline{\eta},\eta}P$, we obtain that
\[
(a,b):=
(\eta+\sqrt E\lambda,\overline{\eta}+\frac{\sqrt E}{\lambda})|_{\eta=\eta_0(\lambda)}=-\sqrt E
(\frac{1}{\lambda},\lambda).
\]
Thus $|a|\neq|b|$ when $\lambda\in\Lambda$, and (\ref{awas1}) is proved. By reducing the constants $\gamma,\rho$ if needed, one may assume that the constants in (\ref{awas}), (\ref{awas1}) coincide.

Let us show that (\ref{dddd}) is equivalent to the following estimate:
\begin{equation}\label{ddddab}
|g(z,k(\lambda)) | \leq Cf(|z|)  \quad {\rm when} \quad z \in \mathbb C, ~~|k|=1 ,
\end{equation}
where $|k|=1$ is equivalent to $E=1/2, ~\lambda \in \Lambda$. Indeed, (\ref{dddd}) can be replaced by
\begin{equation}\label{dddda}
|g(z,k(\lambda)) | \leq Cf(|z|\sqrt {E(|\lambda|^2+|1\backslash \lambda|^2)/2}), \quad z \in \mathbb C, \lambda \in \mathbb C \backslash \{0\}.
\end{equation}
We note that
\[
|k_1|^2=\frac{E}{4}(\lambda+\frac{1}{\lambda})(\overline{\lambda}+\frac{1}{\overline{\lambda}})=\frac{E}{4}(|\lambda|^2+\frac{1}{|\lambda^2|}+2
\Re \frac{\overline{\lambda}}{\lambda}), ~~|k_2|^2=\frac{E}{4}(|\lambda|^2+\frac{1}{|\lambda|^2}-2\Re \frac{\overline{\lambda}}{\lambda}).
\]
Thus
\[
|k|^2=\frac{E}{2}(|\lambda|^2+\frac{1}{|\lambda|^2}),
\]
and the inequality in  (\ref{dddda}) takes the form $|g(z,k) | \leq Cf(|z||k|)$. The substitution $\xi\to |k|\xi$ in the integral (\ref{2DecA}) implies that $g(z,k)=g(z|k|,k/|k|). $ Thus it is enough to prove (\ref{dddda}) when $|k|=1$, i.e., (\ref{dddd}) is equivalent to (\ref{ddddab}).

In order to prove (\ref{ddddab}), we introduce the following cut-off functions: $\alpha\in C^\infty(\mathbb R^2),~ \alpha(\xi)=1$ when $|\xi|<\rho/2,~ \alpha(\xi)=0$ when $|\xi|>\rho,~\alpha_1(\xi)=\alpha(\xi-\xi^{(0)})$, where $\xi^{(0)}_1+i\xi^{(0)}_2=\eta_0(\lambda)$, and $\beta(\xi)=1-\alpha(\xi)-\alpha_1(\xi)$. We represent $g(z,k)$ in the form
\[
g(z,k)=-F^{-1}\frac{1}{P(\xi,\lambda)}=-F^{-1}\frac{\beta(\xi)}{P(\xi,\lambda)}-F^{-1}\frac{\alpha(\xi)}{P(\xi,\lambda)}-
F^{-1}\frac{\alpha_1(\xi)}{P(\xi,\lambda)}:=g_1+g_2+g_3,
\]
where $F^{-1}$ is the inverse Fourier transform.  Recall that we can we assume that $E=1/2,~\lambda\in \Lambda$. Under this assumption, we will estimate each of the terms above.

From (\ref{awas}), (\ref{awas1}) it follows that the distance between points $\xi=0$ and $\xi=\xi^{(0)}$ exceeds $\rho$. Hence $\beta(\xi)=0$ in a neighborhood of points $0,\xi^{(0)}$, i.e., function $\beta(\xi)(\frac{1}{P(\xi,\lambda)}-\frac{1}{|\xi|^2+1})$ and all its derivatives are integrable. Thus for each $N$,
\[
|F^{-1}[\beta(\xi)(\frac{1}{P(\xi,\lambda)}-\frac{1}{|\xi|^2+1})]|\leq \frac{C_N}{1+|x|^N}.
\]
The same estimate holds for $F^{-1}\frac{1-\beta(\xi)}{|\xi|^2+1}$.
Since  $F^{-1}\frac{1}{|\xi|^2+1}$ decays exponentially at infinity and has a logarithmic singularity at the origin, it follows that $|g_1|\leq Cf(|z|)$.

From (\ref{awas}) and the same estimate on the linear approximation $P_1$ of $P$, it follows that $\alpha(\xi)(\frac{1}{P(\xi,\lambda)}-\frac{1}{P_1(\xi,\lambda)})$ and its derivatives of the first order are integrable. Thus
$$
|F^{-1}\alpha(\xi)(\frac{1}{P(\xi,\lambda)}-\frac{1}{P_1(\xi,\lambda)})|\leq \frac{C}{1+|x|}.
$$
The same estimate is valid for $F^{-1}\frac{\alpha(\xi)}{P_1(\xi,\lambda)}$. Indeed, the boundedness follows from the integrability of $\frac{\alpha(\xi)}{P_1(\xi,\lambda)}$. The decay at infinity is the consequence of the following two facts: $F^{-1}\frac{1}{P_1(\xi,\lambda)}$ is a homogeneous function of order $-1$, and $F^{-1}\frac{1-\alpha(\xi)}{P_1(\xi,\lambda)}$ decays at infinity due to the relation \[
F^{-1}\frac{1-\alpha(\xi)}{P_1(\xi,\lambda)}=-\frac{1}{|x|^2}F^{-1}[\Delta\frac{1-\alpha(\xi)}{P_1(\xi,\lambda)}].
\]
Hence $|g_2|\leq \frac{C}{1+|x|}$. Obviously, a similar estimate is valid for $g_3$. Estimates on $g_i$ imply~(\ref{ddddab}).
\qed

\label{010716A}
{\bf II. Proof of (\ref{12NovD}).} To make calculations more transparent, we will justify  (\ref{12NovD}) in the case of $E=1$. Since $\frac{1}{\pi k_j}$ is a fundamental solution for the operator $\partial/\partial \overline{k_j}$, from (\ref{2DecA}) it follows that
\[
\frac{\partial}{\partial \overline{k_j}}g({x},k)=- \frac{1}{2\pi}
\int_{ \mathbb R^2} \xi_j\delta(|\xi|^2 + 2k \cdot \xi)e^{i\xi x }d \xi, \quad  k \in \mathbb C^2, ~\Im k \neq 0.
\]
From (\ref{kla}) it follows that
\[
\frac{\partial}{\partial \overline\lambda}=\frac{1}{2}\left [\frac{\partial}{\partial \overline{k_1}}\left (1-\frac{1}{\overline\lambda^2}\right )+i\frac{\partial}{\partial \overline{k_2}}\left (1+\frac{1}{\overline\lambda^2}\right ) \right ]
\]
and that $|\xi|^2 + 2k \cdot \xi=|\eta|^2 + \lambda \overline{\eta}+ \frac{\eta}{\lambda},~\eta=\xi_1+i\xi_2\in\mathbb C.$ Hence
\[
\frac{\partial}{\partial \overline{\lambda}}g({x},k(\lambda))=- \frac{1}{4\pi}
\int_{ \mathbb C} \left (\eta-\frac{1}{\overline\lambda^2}\overline{\eta}\right )\delta\left (|\eta|^2 + \lambda \overline{\eta}+ \frac{\eta}{\lambda}\right )e^{i\Re(\eta \overline z)} d \xi_1d\xi_2, \quad \eta=\xi_1+i\xi_2 \in \mathbb C.
\]

The points $\eta=\eta(\lambda)$ where the argument of the delta-function vanishes were found in the previous part of the Appendix.
These points are $\eta=0$ and $\eta=\eta_0(\lambda)=-\lambda-\frac{1}{\overline{\lambda}}$. The point $\eta=0$ does not contribute to the integral because the first factor in the integrand vanishes at this point. Thus
\begin{equation}\label{ggg}
\frac{\partial}{\partial \overline{\lambda}}g({x},k(\lambda))=- \left [\frac{ 1}{4\pi}\left (\eta-\frac{1}{\overline\lambda^2}\overline{\eta} \right )|J(\eta)|^{-1}e^{i\Re(\eta \overline z)}\right ]_{\eta=\eta_0(\lambda)}, \quad  \eta=\xi_1+i\xi_2\in \mathbb C,
\end{equation}
where $J$ is the Jacobian of the real and imaginary parts of $f(\eta):=|\eta|^2 + \lambda \overline{\eta}+\frac{\eta}{\lambda}$ with respect to $\xi_1,\xi_2$.

In order to evaluate the Jacobian, we make a shift $\eta=-\lambda-\frac{1}{\overline{\lambda}}+u, ~u=u_1+iu_2,$ in the argument of $f$. This leads to
\[
f=-\overline{\lambda}u-\frac{1}{\overline{\lambda}}\overline{u}+|u|^2.
\]
Thus
\[
f_1:=\left . f'_{u_1} \right |_{u=0}=-\overline{\lambda}-\frac{1}{\overline{\lambda}},~~f_2:=\left . f'_{u_2}\right |_{u=0}=\left (-\overline{\lambda}+\frac{1}{\overline{\lambda}} \right )i,
\]
and
\[
J=\left |\det\left(
               \begin{array}{cc}
                 \Re f_1 & \Im f_1 \\
                 \Re f_2 & \Im f_2 \\
               \end{array}
             \right) \right |=\left |\Im(\overline{f_1}f_2)\right |=\left |\Re \left [\left (\lambda+\frac{1}{\lambda} \right )\left (\overline{\lambda}-
             \frac{1}{\overline{\lambda}} \right ) \right ]\right |=|\lambda|^2-\frac{1}{|\lambda|^2}
=\frac{|1-|\lambda|^4|}{|\lambda|^2}.
\]
It is easy to check that
\[
\left . \left (\eta-\frac{1}{\overline\lambda^2}\overline{\eta} \right ) \right |_{\eta=-\lambda-1/\overline{\lambda}}=
\frac{1-|\lambda|^4}{\overline{\lambda}|\lambda|^2}.
\]
This, together with the previous formula, (\ref{ggg}), and relations (\ref{2DecA}), (\ref{exp}) between $G$ and $g$ completes the proof of (\ref{12NovD}).
\qed

{\bf III. Derivation of the $\partial$-bar equations (\ref{dbar}), (\ref{dbar1}).}
Let us derive $\overline{\partial}$-equation (\ref{dbar}) following the arguments of Grinevich-Manakov \cite{gm}.
We will write the Lippman-Schwinger equation (\ref{19JanA}) in the form
\begin{equation}\label{15DecE1}
(I-G(\lambda) v)\psi(z,\lambda)=  e^{i\frac{\sqrt{E}}{2} (\lambda \overline{z} + z / \lambda) },
\end{equation}
where $v$ is the operator of multiplication by the potential $v$ and $G(\lambda)$ is the  operator of convolution with the Green function $G({x},k)$.
After differentiating  (\ref{15DecE1}) in $\overline{\lambda}$ and taking into account (\ref{12NovD}) and (\ref{27dec1}), we get that
\begin{equation}\label{15DecE2}
(I-G(\lambda)v)\frac{\partial \psi}{\partial \overline{\lambda}}=  \frac{\partial G}{\partial \overline{\lambda}} v \psi=\frac{\sgn (|\lambda|^2-1)}{4\pi \overline{\lambda}}
e^{-i\sqrt{E}/2 (\overline{\lambda} z + \overline{z} / \overline{\lambda})}*(v\psi)=r(\lambda)e^{-i\sqrt{E}/2 (\overline{\lambda} z + \overline{z} / \overline{\lambda})},
\end{equation}
where $|\lambda|\neq 0,1$. Thus
\begin{equation} \label{hhhj}
\frac{\partial \psi}{\partial \overline{\lambda}}=r(\lambda)(I-G(\lambda)v)^{-1}e^{-i\sqrt{E}/2 (\overline{\lambda} z + \overline{z} / \overline{\lambda})},
\end{equation}
where $u=(I-G(\lambda)v)^{-1}f$ is understood in the following sense: we solve the equation $(I-G(\lambda)v)u=f$ on the compact ${\rm supp} v$, and then define $u$ for all $z$ as $u = f + G(\lambda)( v u)$. Since $G(\lambda)=G(-\frac{1}{\overline{\lambda}})$ (see \cite[(27)]{gm},\cite[lemma 3.1]{g2000} ), from  (\ref{15DecE1}) it follows that the right-hand side in (\ref{hhhj}) is equal to the right hand side in (\ref{dbar}). Since the Green function $G$ is real valued, from (\ref{15DecE1}) it follows that the right-hand side in the last equation in the case of real valued potential $v$ coincides with the right-hand side in (\ref{dbar1}).
\qed

{\bf IV. Compactness of $\overline{\partial}^{\!~-1}(g \cdot).$}
Recall that operator $\overline{\partial}^{\!~-1}$ is defined in (\ref{dmin}). We recall the proof (see \cite[Lemma 3.1]{perry} or \cite[Lemma 5.3]{music}) of the following fact.
\begin{lemma}
Let $g \in L^2$. Then the operator $T_g:u \rightarrow \overline{\partial}^{\!~-1}(gu)$ is compact in $L^p(\mathbb C), ~p>2$.
\end{lemma}
{\bf Proof.} The Hardy-Littlewood-Sobolev inequality implies that
\begin{equation}\label{2007A}
\| \overline{\partial}^{\!~-1}f\|_{L^{\widetilde{q}}} \leq C_q \|f\|_{L^q}, \mbox{ where } \frac{1}{\widetilde{q}}=\frac{1}{q}-\frac{1}{2}.
\end{equation}
Let $\widetilde{q}=p>2$ and $q=\frac{2p}{p+2}$. From (\ref{2007A}) and the H$\ddot{o}$lder inequality, it follows that
\begin{equation}\label{2007B}
\| \overline{\partial}^{\!~-1}(gu)\|_{L^{p}} \leq C_p \|g\|_{L^2} \|u\|_{L^p}.
\end{equation}
The Beurling transform is defined on $C_0^\infty(\mathbb C)$ by
$$
(Sf)(\lambda)= - \frac{1}{\pi} \lim_{\varepsilon \downarrow 0 } \int_{|\lambda-\varsigma|>\varepsilon}\frac{d\varsigma_R d\varsigma_I}{(\lambda-\varsigma)^2} f(\varsigma).
$$
We will need
\begin{lemma}\label{2007C}(\cite[4.3]{astala}).
The operator $S$ extends to a bounded operator from $L^p(\mathbb C)$ to $L^p(\mathbb C)$ for each $p \in (1,\infty)$, and as unitary operator if $p=2$.  Moreover, if $\nabla \varphi \in L^q$ for some $q>1$, then $S(\overline{\partial} \varphi)=\partial \varphi$.
\end{lemma}

By the norm-closedness of compact operators, the estimate (\ref{2007B}), and the density of $C_0^\infty(\mathbb C)$ in $L^2(\mathbb C)$, it suffices to show
that $T_g$ is compact for $g \in C_0^\infty$. Let $\Omega \in \mathbb C$ be the  ball containing the support of $g$. Let $p'\in (1,2)$ be the conjugate exponent to $p$. It suffices to show that the adjoint operator $T_g^*$ is compact in $L^{p'}$. If $f \in L^{p'}(\mathbb C)$, then $\overline{\partial}^{\!~-1}f \in L^{\frac{2p}{p-2}}(\mathbb C)$ by inequality (\ref{2007A}), while $\nabla \overline{\partial}^{\!~-1} f \in L^{p'}(\mathbb C)$ by lemma  \ref{2007C}. Thus
$$
\|g\overline{\partial}^{\!~-1}f\|_{W^{1,p'}} \leq C \left ( 1+|\Omega|^{1/2} \right ) \|f\|_{L^{p'}},
$$
and the compactness follows from the Rellich-Kondrachov embedding theorem.
\qed


{\bf V. Dirichlet-to-Neumann map.} Consider the problem
\begin{equation}\label{dir3}
(\Delta+E-v)u(x) =0,~~~  x \in \mathcal O; \quad u=\phi, ~~~ x\in \partial\mathcal O,
\end{equation}
where $v\in L^p(\mathcal O), ~u\in W^{2,p}(\mathcal O),~u_0\in W^{2-1/p,p}(\partial\mathcal O),~p>1$ (see the definitions of above spaces in,  e.g., \cite{agdn}, \cite{agranovich}). The Dirichlet-to-Neumann map $\Lambda_v$ is defined as the operator
$$
\Lambda_v: W^{2-1/p,p}(\partial\mathcal O)\to W^{1-1/p,p}(\partial\mathcal O), \quad \Lambda_vu_0=\frac{\partial u}{\partial \nu}|_{\partial\mathcal O}.
$$
The unique solvability of (\ref{dir3}) is needed in order for $\Lambda_v$ to exist.

It was assumed in the introduction that $E$ was not an eigenvalue of the operator $-\Delta+v$. This is understood in the sense that the homogeneous problem (\ref{dir3}) has only the zero solution. However, the solvability of  (\ref{dir3}) does not follow immediately from the uniqueness since the standard proofs of the Fredholm property for elliptic problems require certain smoothness of the coefficients, and we need one more step to justify the solvability when $v\in L^p(\mathcal O)$.

\begin{lemma}
If $E$ is not an eigenvalue of $-\Delta+v$, then the solution of the problem  (\ref{dir3}) exists, and operator $\Lambda_v$ is well defined.
\end{lemma}
{\bf Proof.} Let $w_1=Bu_0$ and $w_2=\Delta^{-1}g$ be $W^{2,p}$-solutions of the problems
\begin{equation}\nonumber
\Delta w_1(x) =0,~~  x \in \mathcal O, ~~w_1|_{\partial\mathcal O}=u_0; \quad~~ \Delta w_2(x) =g\in L^p(\mathcal O),~~  x \in \mathcal O,~~ w_2|_{\partial\mathcal O}=0.
\end{equation}
These problems are uniquely solvable \cite{agdn}, and the operators
\[
B: ~W^{2-1/p,p}(\partial\mathcal O)\to W^{2,p}(\mathcal O), ~~~\Delta^{-1}:~L^{p}(\mathcal O)\to W^{2,p}(\mathcal O)
\] are bounded. We look for the solution of (\ref{dir3}) in the form
\begin{equation}\label{dir6}
u=Bu_0+\Delta^{-1}g,
\end{equation}
 and obtain the following equation for $g$:
\begin{equation}\label{dir5}
g+(E-v)\Delta^{-1}g=f, \quad {\rm where }~~~f=(-E+v)Bu_0.
\end{equation}
Moreover, formula  (\ref{dir6}) establishes the one-to-one correspondence between solutions $u$ of  (\ref{dir3}) and solutions $g\in L^p$ of  (\ref{dir5}). Indeed, it is easy to see that function $u$ given by  (\ref{dir6}) with $g$ satisfying (\ref{dir5}) is a solution of  (\ref{dir3}). Conversely, let $u$ be a solution of  (\ref{dir3}). We define $g=(E-v)u$. Then $u$ can be written in the form of (\ref{dir6}), and therefore $g$ satisfies (\ref{dir5}).

From the Sobolev inequalities it follows that the embedding $W^{2,p}(\mathcal O)\to C(\mathcal O)$ is a compact operator (since the embedding into $C^\alpha(\mathcal O)$ is bounded when $n=2$ and  $\alpha>0$ is small enough). From here it follows that $f\in L^p$ and that the operator $(E-v)\Delta^{-1}:L^p\to L^p$ is compact. Thus equation (\ref{dir5}) is Fredholm. The uniqueness of its solution follows from the uniqueness for   (\ref{dir3}), and the existence of the solution $g$ of  (\ref{dir5}) implies the existence of the solution $u$ of (\ref{dir3}).

\qed

{\bf Acknowledgments.} This paper includes many ideas from our joint paper with R.G. Novikov \cite{NLV}. The authors are also  grateful to Keith Rogers, Daniel Faraco Hurtado and Jorge Tejero Tabernero  for productive discussions.

\end{document}